\documentclass[12pt]{article}

\usepackage{graphicx,epsfig}

\usepackage{amsmath}
\usepackage{amsfonts}

\newcommand{\beq}{\begin{eqnarray}}
\newcommand{\eeq}{\end{eqnarray}}

\def\ni{\noindent}
\def\ms{\medskip}

\def\nn{\nonumber}
\def\bce{\begin{center}}
\def\ece{\end{center}}

\begin{document}

\vspace*{-10mm}




\thispagestyle{empty}

\vspace*{4mm}

\begin{center}

{\large \bf  Stationary vs.~singular points in an accelerating
FRW cosmology \\
derived from six-dimensional Einstein-Gauss-Bonnet gravity}

\vspace{4mm}

\medskip

{\sc E. Elizalde}$^{1,}$\footnote{Presently on leave at
Department of Physics \& Astronomy,
Dartmouth College, 6127 Wilder Laboratory, Hanover, NH 03755, USA.
E-mail: elizalde@ieec.uab.es,
elizalde@math.mit.edu},
{\sc A.N. Makarenko$^{2,}$\footnote{E-mail: andre@tspu.edu.ru}, V.V.
Obukhov$^{2,}$\footnote{E-mail: obukhov@tspu.edu.ru},
\\ K.E. Osetrin$^{2,}$\footnote{E-mail: osetrin@tspu.edu.ru},
and A.E. Filippov}$^{2}$

\vspace{3mm}

$^{1}${\sl Instituto de Ciencias del Espacio (CSIC) and Institut
d'Estudis Espacials de Catalunya (IEEC/CSIC), Facultat de
Ci\`encies,  Universitat Aut\`onoma de Barcelona \\
 Torre C5-Parell-2a planta, 08193 Bellaterra
(Barcelona) Spain} \\ \vspace*{2mm}

$^{2}${\sl Laboratory for Fundamental Study, Tomsk State Pedagogical
University \\ 634041 Tomsk, Russia}

\vspace{16mm}

{\bf Abstract}

\end{center}

Six-dimensional Einstein-Gauss-Bonnet gravity (with a linear
Gauss-Bonnet term) is investigated. This theory is
 inspired by basic features of results coming from string and M-theory.
Dynamical compactification is carried out and it is seen that a
four-dimensional accelerating FRW universe is recovered,
 when the two-dimensional internal space radius shrinks.
A non-perturbative structure of the corresponding theory is
identified which has either three or one stable fixed points,
depending on the Gauss-Bonnet coupling being positive or negative. A
much richer structure than in the case of the perturbative regime of
the dynamical compactification recently studied by Andrew, Bolen,
and Middleton is exhibited.

\vspace{3mm}


\vspace{2mm}


\newpage

\ni {\bf 1. Introduction.}
To realize that the expansion of the universe is accelerating was
one of the most important scientific discoveries of the last
century. There are several alternative explanations of this
remarkable fact (what already means that it is not well understood
yet). A quite appealing possibility for the gravitational origin of
the dark energy responsible for this accelerated expansion is the
modification of General Relativity (GR) or the corresponding
Einsteinian gravity (see \cite{no06a} for a review of these
approaches). Such a well-established and successful theory cannot be
modified without a very good reason, and much less in an arbitrary
way. But observe that there is no compelling reason why standard GR
should be trusted at cosmological scales. For a rather minimal
modification of the same, one assumes that the gravitational action
might contain some additional terms which would start to grow slowly
with decreasing curvature. A particularly interesting formulation is
obtained by using well-grounded geometrical arguments, specifically
when the modified gravity action is endowed with a function of the
Gauss-Bonnet (GB) topological invariant, $G$, as it was suggested in
\cite{205a}.

It must be noted that different types of dark energy may actually
show up in different ways, at large distances. Cold dark matter is
known to be localized near galaxy clusters but, quite on the
contrary, dark energy distributes uniformly in the universe. The
reason for that behavior could be explained by a difference in the
equation of state parameter $w=p/\rho$. Moreover, the effect of
gravity on the cosmological fluid turns out to depend on $w$ and it
so happens that, even when $-1<w<0$, gravity can act sometimes as a
repulsive force. The effect of gravity on matter with $-1<w<0$ can
be shown to be opposite to that on usual matter, which becomes dense
near a star, while matter with $-1<w<0$ becomes less dense when
approaching a star \cite{205a}. Dark energy contributes uniformly
throughout the universe, which would be indeed consistent, since the
equation of state parameter of dark energy is almost $-1$. If dark
energy is of phantom nature ($w<-1$), its density becomes large near
the cluster but if dark energy is of quintessence type
($-1<w<-1/3$), then its density becomes smaller.

Another very important argument in favor of the GB modified theory
is the fact that it can be seen as being inspired by string and/or
M-theory \cite{no03a}. In fact, specific models for string-inspired
scalar GB gravity, considered as possible forms of dark energy, have
been discussed in \cite{no05a} and \cite{sttt}. It was subsequently
shown in \cite{nos06a}, that scalar GB gravity can actually be
represented as a modified GB gravity without scalars, and
specifically that it can be equivalent to an ideal fluid with an
homogeneous equation of state \cite{no05b}.

In the present paper we will study the case of 6-dimensional
Einstein GB gravity, with a linear GB term. As already advanced,
this situation is quite interesting, since the theory that we will
thereby obtain can be shown to be inspired by what is derived from
string/M-theory, after some specific compactification is carried
out, when the scale factor of the 2-dimensional internal space goes
to zero. In the course of our study we will discover a
non-perturbative structure of the corresponding theory with either
three (for positive GB coupling, $\epsilon$) or one (for
negative GB coupling)
  stable fixed points. This will exhibit a much richer
structure than the one that follows from the case of the
perturbative regime of the dynamical compactification recently
studied by Andrew et al. \cite{a06a} \ms

\ni {\bf 2. Dynamical compactification of 6-dimensional
Einstein-Gauss-Bonnet gravity.}
We shall start from the following, string-inspired action in six
dimensions
 \beq S=\int d^6x\sqrt{-g}(R+\epsilon L_{GB}),\eeq
where $\epsilon$ is a constant, and the metric is the product of the
usual metric corresponding to the 4-dimensional FRW universe and a
2-dimensional surface, namely \beq
ds^2=-dt^2+a^2(t)[(dx^1)^2+(dx^2)^2+(dx^3)^2]+b^2(t)[(dx^4)^2+(dx^5)^2],\eeq
the scalar curvature is \beq R=\frac{6{\dot a}^2}{a^2}+\frac{12\dot
a\dot b}{ab}+\frac{2{\dot b}^2}{b^2}+\frac{6\ddot a}a+\frac{4\ddot
b}b,\eeq
 while the four-dimensional and topologically invariant Gauss-Bonnet
Lagrangian, $L_{GB}$, has the form
\beq L_{GB}=\frac{48{\dot a}^3\dot b}{a^3b}+\frac{72{\dot a}^2{\dot
b}^2}{a^2b^2}+\frac{24{\dot a}^2\ddot a}{a^3}+
    \frac{96\dot a\ddot a\dot b}{a^2b}+\frac{24\ddot a{\dot b}^2}{ab^2}
    +\frac{48{\dot a}^2\ddot b}{a^2b}+
    \frac{48\dot a\dot b\ddot b}{ab^2},\eeq
or, equivalently, \beq L_{GB}=\frac{24}{a^3b^2}(2{\dot a}^3b\dot
b+3a{\dot a}^2{\dot b}^2+{\dot a}^2\ddot ab^2+4a\dot a\ddot ab\dot
b+
    a^2\ddot a{\dot b}^2+2a{\dot a}^2b\ddot b+2a^2\dot a\dot b\ddot b).\eeq
Note that, unlike the six-dimensional compatification case of
Elizalde et al. \cite{eko1}, the theory under consideration is
not multiplicatively renormalizable.
The corresponding equations of motion are obtained by variation of
the action with respect to $a$ and $b$, what yields \beq {\dot
a}^2b^2+4a\dot ab\dot b+a^2{\dot b}^2+2a\ddot ab^2+2a^2b\ddot
b+12\epsilon {\dot a}^2{\dot b}^2+
    16\epsilon \dot a\ddot ab\dot b+8\epsilon a\ddot a{\dot b}^2 &&
\nn \\
    +8\epsilon {\dot a}^2b\ddot b+16\epsilon a\dot a\dot b\ddot b&=&0,\nn \\
     3a{\dot a}^2b+3a^2\dot a\dot b+3a^2\ddot ab+a^3\ddot b+12\epsilon
{\dot a}^3\dot b+12\epsilon {\dot a}^2\ddot ab+
    24\epsilon a\dot a\ddot a\dot b+12\epsilon a{\dot a}^2\ddot b&=&0.\label{2a}\eeq
These equations can be easily rewritten in terms of the Hubble rates
 \ $H=a'/a$ and $h=b'/b$, namely
\beq 3h^2+4hH+3H^2+2\dot h+2\dot H+16\epsilon h^3H+28\epsilon
h^2H^2+16\epsilon hH^3 &&
\nn \\ +16\epsilon h\dot h H+8\epsilon \dot h H^2
+8\epsilon h^2\dot H+16\epsilon hH\dot H&=&0,\nn \\
 h^2+3hH+6H^2+\dot h+3\dot H+12\epsilon h^2H^2+36\epsilon
hH^3+12\epsilon H^4 && \nn \\ +12\epsilon \dot h H^2+24\epsilon
hH\dot H+12\epsilon H^2\dot H&=&0.\label{2b}\eeq In addition,
variation over the metric in the above expressions gives the
constraint equation \beq h^2+6hH+3H^2+36\epsilon h^2H^2+24\epsilon
hH^3=0 \label{connect} \eeq This equation helps to exclude $h$ and
$h'$ from Eqs. (\ref{2b}). As a result, one gets an equation for $H$
only: \beq && \hspace*{-8mm} H'= 3H^2 \times \\ && \hspace*{-5mm}
\frac{\sqrt 6+4G+\epsilon (-22\sqrt
6+64G)H^2\hspace*{-1.5mm}-\hspace*{-1mm}24\epsilon ^2(9\sqrt
6-52G)H^4\hspace*{-1.5mm}+\hspace*{-1mm}96\epsilon ^3(17\sqrt
6+12G)H^6\hspace*{-1.5mm}-\hspace*{-1mm}8064\sqrt 6\epsilon ^4H^8}
{\sqrt 6-12\epsilon (\sqrt 6+16G)H^2+72\epsilon ^2(3\sqrt6
-32G)H^4-2880\sqrt 6\epsilon ^3H^6+31104\sqrt 6\epsilon ^4H^8},\nn
\eeq where \beq G=\sqrt{1-6\epsilon H^2+24 \epsilon^2H^4}.\eeq One
can check that this last equation obeys the fundamental relation
(for $\epsilon >0$):
\beq H'=\frac{H^2(H^2-p^2)}{(H^2-q^2)(H^2-r^2)}f(H),\eeq where $p,
q$ and $r$ are constants, and the function \ $f(H)<0$.
We start now with its numerical analysis.
 \ms

\ni {\bf 3. First case: $\epsilon >0$.}
It leads to the following
values of the constants, corresponding to constant curvatures in the
4-dimensional and 2-dimensional spaces: \beq p_0\approx
0,7501/\sqrt{\epsilon},\quad q_0\approx 0,4842/\sqrt{\epsilon},\quad
r_0\approx 0,1023/\sqrt{\epsilon}. \eeq Note that $p_0$ is a stable
fixed point while $q_0$ and $r_0$ are singular points.
It turns out then, that the initial values of $H(t)$ can be
classified as belonging to four different regions, which are
delimited by these values of $p_0, q_0$ and $r_0$. The behavior of
$H(t)$ in each of these regions can differ considerably, from one to
the other. We will now consider the different possibilities in
detail.
The four different cases corresponding to $\epsilon >0$ are as
follows. \ms

\ni 1. Case $0<H(0)<r_0$, then $H'>0$, when $H\to q_0$, and in the
limit it turns out that $H'\to +\infty$. Thus, at $H=q_0$ a
singularity develops, Figs.~1a, 1b  (for a classification of the
different types of future singularities, see e.g. \cite{not05a}):

\includegraphics[116,0][100,150]{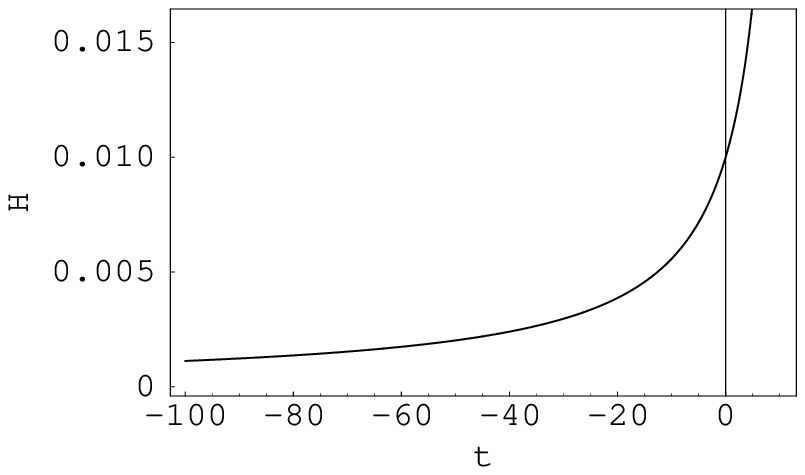}
\includegraphics[-150,0][100,100]{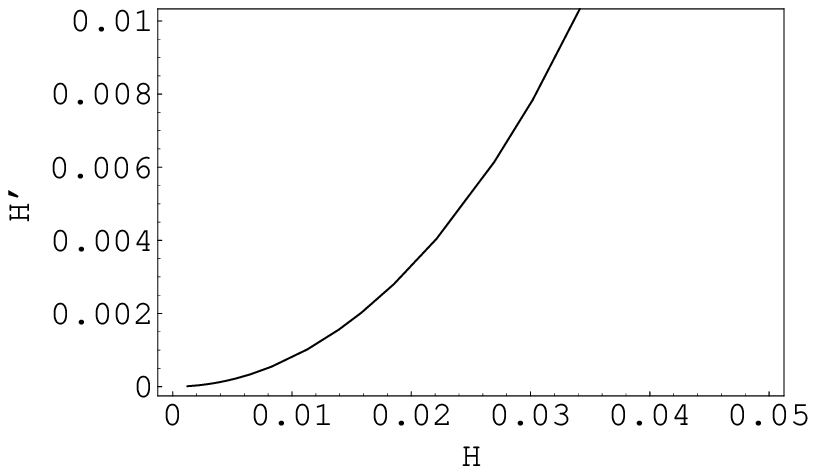}
\hspace*{-6.5cm} Fig.~1a \hspace*{7.5cm} Fig.~1b \ms

\ni Using Eq.~(\ref{connect}), we obtain the form of $h(t)$ and
$h'(h)$, Figs.~1c, 1d (remember that $h=\dot{b}/b$):

\includegraphics[116,0][100,160]{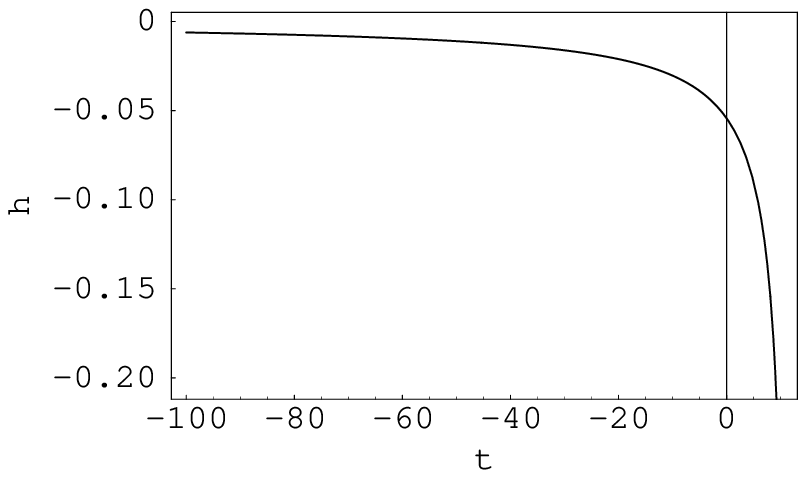}
\includegraphics[-150,0][100,160]{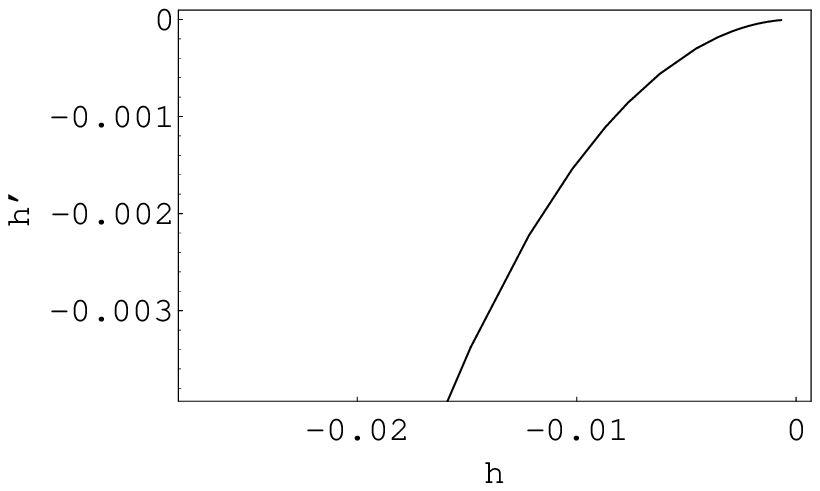}
\hspace*{-6.5cm} Fig.~1c \hspace*{7.5cm} Fig.~1d \ms

\ni We see that in this case the resulting FRW universe expands with
acceleration while the compactification radius of the
extra-dimensional space decreases with time. This indicates that the
higher-dimensional GB term can indeed play the role of dark energy
in this universe. As is the cases in the majority of the dark energy
models currently available, a future singularity occurs. \ms

 \ni 2. Case $r_0 <
H(0) <q_0$, then $H'<0$, when $H\to r_0$, and in the limit one has
that $H'\to -\infty$. Again, $H=r_0$ corresponds to a singularity,
and the curves are, respectively (Figs.~2a, 2b):

\includegraphics[116,0][100,160]{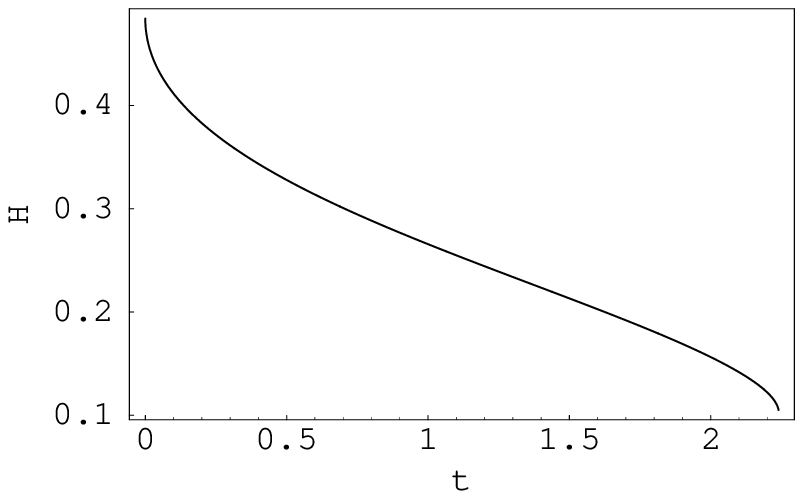}
\includegraphics[-150,0][100,160]{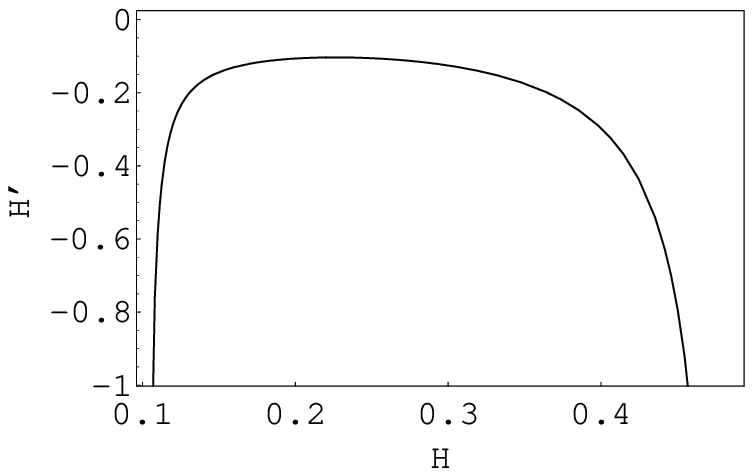}
\hspace*{-6.3cm} Fig.~2a \hspace*{7.2cm} Fig.~2b \ms

\ni As in the case before, using Eq.~(\ref{connect}) the behavior of
$h(t)$ and $h'(h)$ can be obtained (Figs.~2c, 2d)

\includegraphics[116,0][100,160]{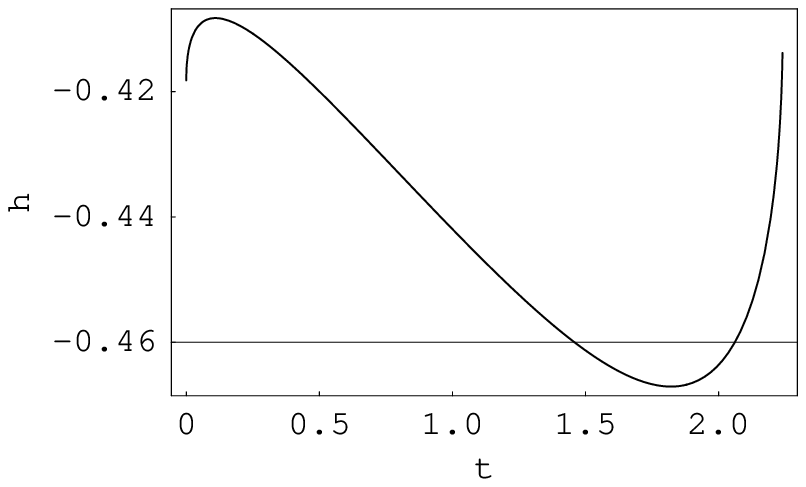}
\includegraphics[-150,0][100,160]{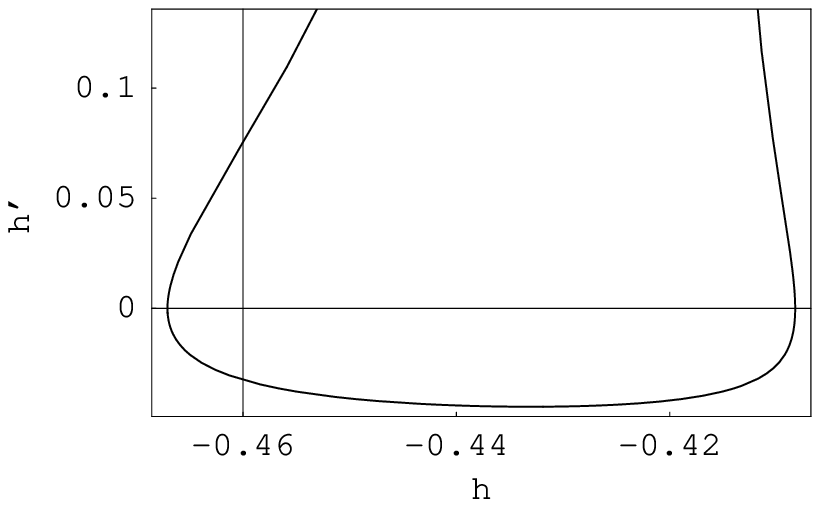}
\hspace*{-6.3cm} Fig.~2c \hspace*{7.2cm} Fig.~2d \ms

\ni We see that in this case the size of the FRW universe decreases,
while the 2-dimensional internal space scale factor may increase.
This case cannot thus naturally describe the dark energy
universe.\ms

\ni 3. Case
 $q_0 <H(0)<p_0$, then $H'>0$, when $H\to p_0$, and then $H$ tends
to a constant, after an oscillatory regime (Figs.~3a, 3b).

\includegraphics[116,13][100,160]{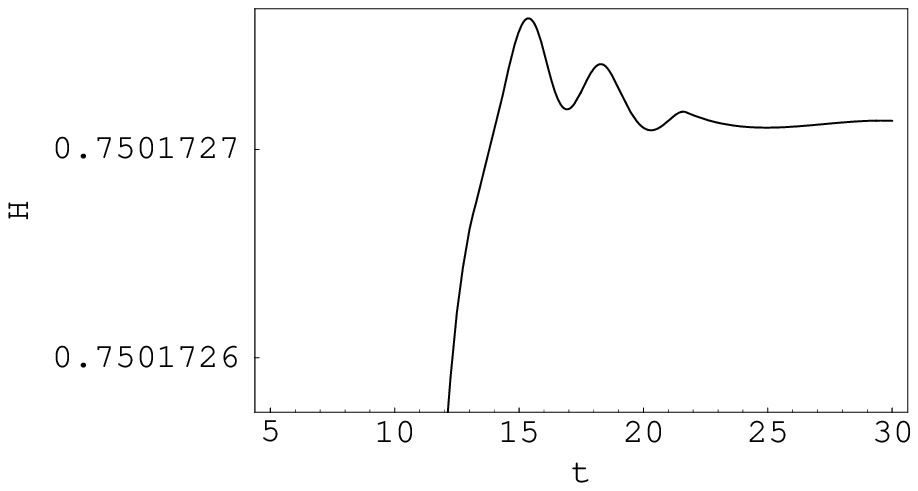}
\includegraphics[-160,0][100,160]{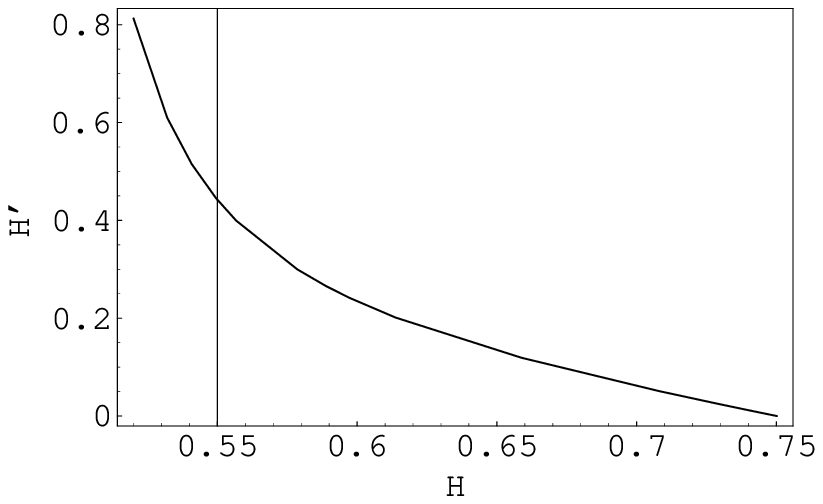}
\hspace*{-5.5cm} Fig.~3a \hspace*{6.5cm} Fig.~3b \ms

\ni Furthermore, the scale factor of the internal space can actually
decrease, in fact (Figs.~3c, 3d)

\includegraphics[116,5][100,160]{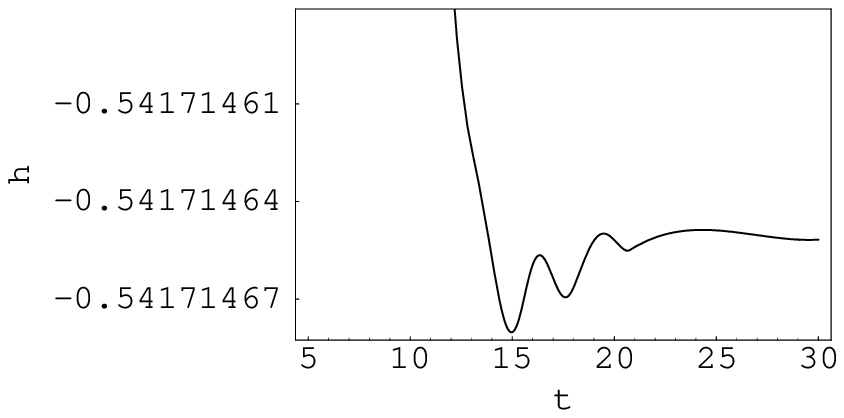}
\includegraphics[-150,0][100,160]{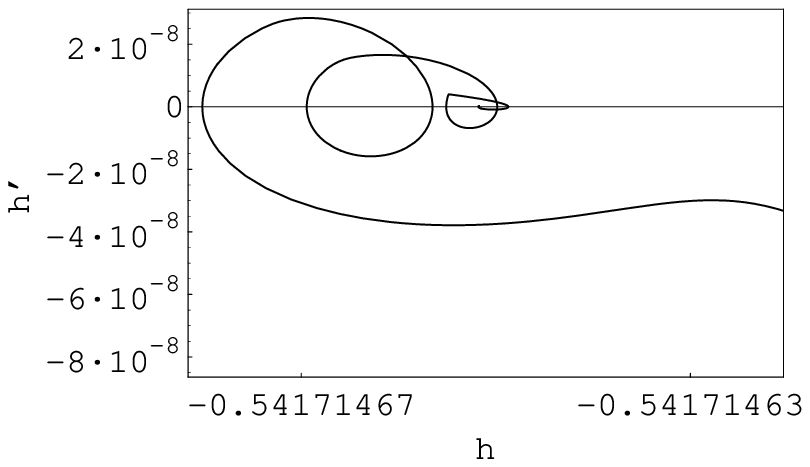}
\hspace*{-5.5cm} Fig.~3c \hspace*{6.5cm} Fig.~3d \ms

\ni Hence, the exact solution obtained at $H=p_0$ turns out to be a
stationary stable point. Precise analysis corresponding to specific
values of the parameters can be further carried out in a rather
simple way.\ms

\ni 4. Case
 $p_0 < H(0)$, then $H'<0$ as $H\to p_0$, and the solution in this
case is oscillatory (Figs.~4a, 4b).

\includegraphics[116,10][100,160]{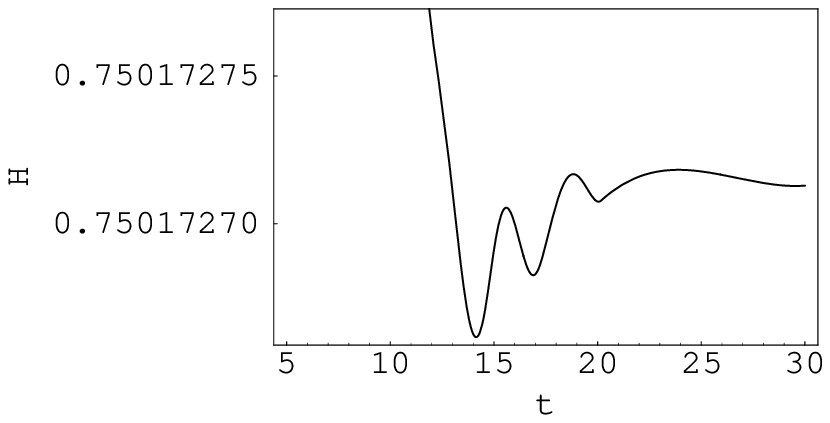}
\includegraphics[-150,0][100,160]{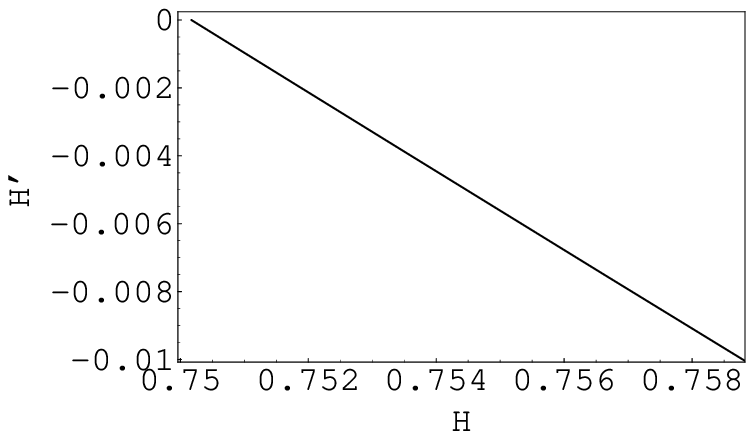}
\hspace*{-5.5cm} Fig.~4a \hspace*{6.5cm} Fig.~4b \ms

\ni As in the previous cases, using (\ref{connect}) the behaviors of
$h(t)$ and $h'(h)$ can be obtained (Figs.~4c, 4d)

\includegraphics[116,3][100,160]{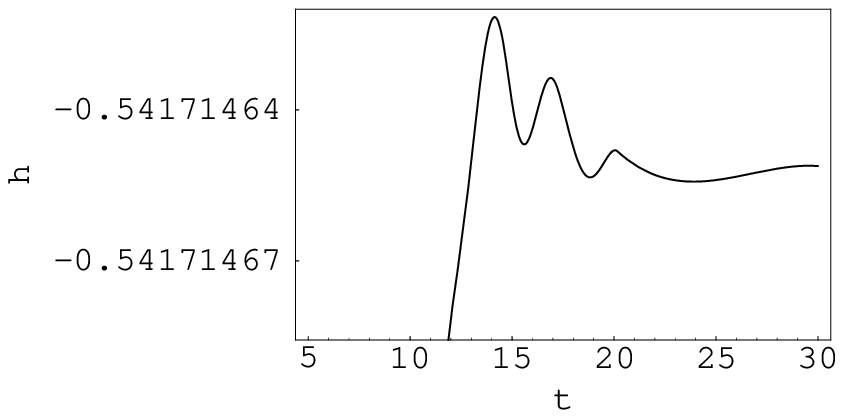}
\includegraphics[-150,0][100,160]{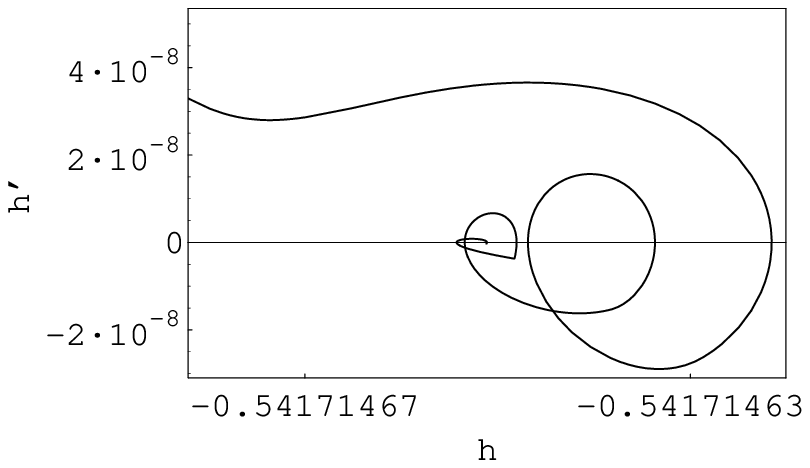}
\hspace*{-5.5cm} Fig.~4c \hspace*{6.5cm} Fig.~4d \ms

\ni In this situation the FRW scale factor tends to a stationary
point with decreasing $H$. The dark energy universe can correspond,
in this case, to one of the branches of the oscillatory universe:
expansion is turned into contraction, and vice-versa, with the
repeated oscillations. Such universe would explicitly correspond to
the general case described in \cite{no0637}. The analytic solution
for this case can be obtained.

Thus, it turns that in the FRW universe the scale factor behavior is
much less complicated for these models than in the case of the
absence of the GB term. This is certainly rewarding. In particular,
for instance, in order to keep close to the Einstein regime, the
condition of proportionality (perturbative regime) of $a(t)$ and
$b(t)$ was used for the study of dynamical compactification in the
Einstein-GB theory in Ref.~\cite{a06a}. For comparison, one can
easily check that the situation there gets quite complicated and
that different regimes for the Hubble rates appear, everything being
much more simple and natural in our model above.
 \ms

\ni {\bf 4. Second case: $\epsilon <0$.}
In this case there is one stationary stable point $H=h=\pm
1/\sqrt{6|\epsilon|}$.

\includegraphics[116,5][100,160]{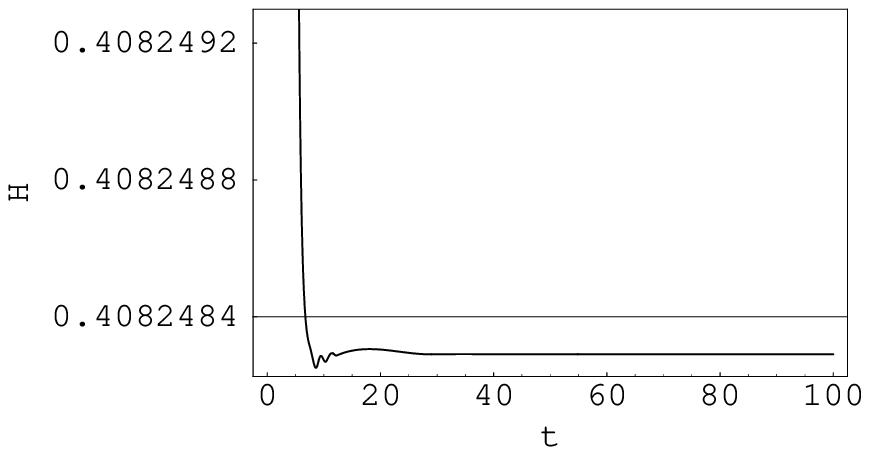}
\includegraphics[-150,0][100,160]{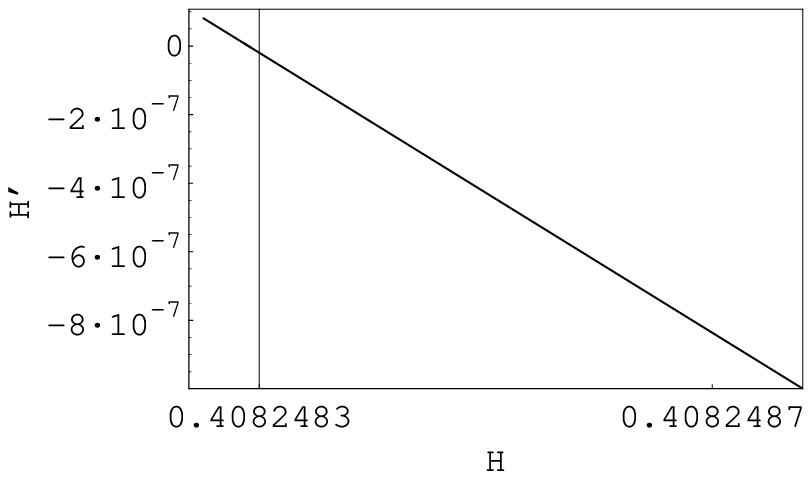}
\hspace*{-5.5cm} Fig.~5a \hspace*{6.5cm} Fig.~5b  \ms

\includegraphics[116,0][100,160]{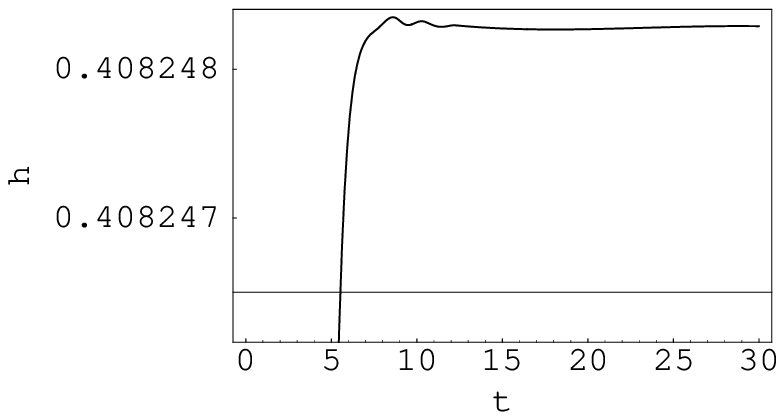}
\includegraphics[-150,0][100,160]{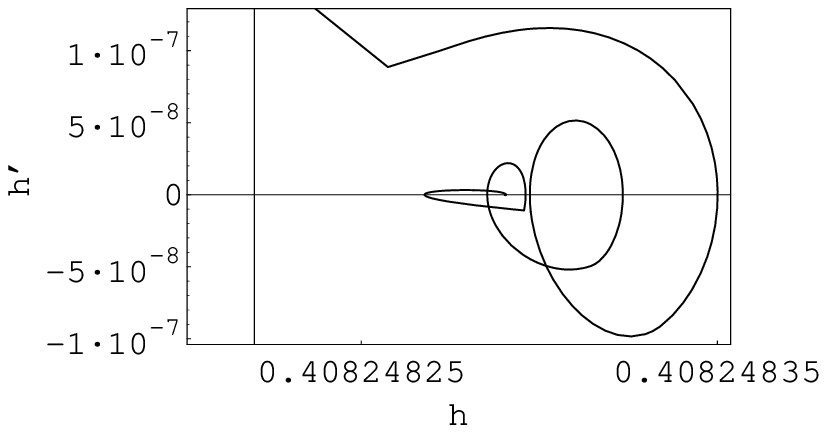}
\hspace*{-6.0cm} Fig.~5c \hspace*{7.1cm} Fig.~5d  \ms

Notice here that solutions with close initial data tend to the
stationary point in the oscillatory regime. However, both scale
factors have the same sign and no dynamical compactification
occurs! The universe, as a whole 6-dimensional object,
either exponentially expands or shrinks. This indicates already that
problems should be expected from a negative$-\epsilon$ model (indeed,
in the higher-dimensional black hole case this may lead to negative
entropy \cite{cve1}).
 \ms

\ni {\bf 5. Third case: $\epsilon =0$.}
In this case one does recover (as it should be)
an explicit solution. In fact, the equations for $a, b$ are
\beq
b^2{a'}^2+4aba'b'+a^2{b'}^2+2ab^2a''+2a^2bb''&=&0,\nn\\
3ab{a'}^2+3a^2a'b'+3a^2ba''+a^3b''&=&0,\eeq and, from here, \beq
3h^2+4hH+3H^2+2h'+2H'&=&0,\nn \\
h^2+3hH+6H^2+h'+3H'&=&0.\eeq From where one gets that \beq
H'=(3\pm 2\sqrt 6)H^2\eeq and the solution is given by \beq H=-\frac
1{\alpha t+C_1},\eeq being \beq \alpha=3\pm 2\sqrt 6.\eeq Moreover, in
terms of the scale factors:
 \beq a&=&C_2[\pm (\alpha t+C_1)]^{-1/\alpha}, \nn \\
b&=&C_3[\pm (\alpha t+C_1)]^{\beta/\alpha},\eeq where
\beq \beta=3\pm \sqrt
6.\eeq

\ni {\bf 6. Discussion.}
Summing up, we have investigated in this paper explicit
non-perturbative dynamical compactification to a 4-dimensional FRW
universe starting from a model of 6-dimensional Einstein-GB gravity.
The number of stationary points obtained depends on the sign of the
GB parameter, $\epsilon$. They correspond in fact to exact solutions
where the curvatures of the 2-dimensional and 4-dimensional spaces
are constant. We have found a regime where the FRW universe does
expand with acceleration, at the same time that the scale factor of
the internal space goes to zero. But the most interesting regime
discovered here is the oscillatory case, where a unification of an
inflationary epoch with a late-time acceleration regime is indeed
possible. This was suggested in \cite{no03147} and has been
explicitly realized here with concrete solutions and numbers, by
combining numerical with analytical methods. For lack of space, in
this letter we have just provided a brief presentation of the
results. A more detailed study of the different accelerating
universes that are derived from the model will be given elsewhere,
in particular, details on the sequence of the matter dominated
phase, the transition to acceleration, and the proper accelerating
phases, as was done for other dark energy models of modified gravity
in \cite{nost}. It will be also necessary to include in our
considerations the effects of scalars (dilaton, moduli) in the
low-energy sector of the string effective action.

\vspace*{2mm}

\ni {\bf Acknowledgments.} We thank S. Nojiri and S.D. Odintsov for
important discussions and correspondence. This work was supported in
part by MEC (Spain), projects BFM2003-00620 and PR2006-0145,  by
AGAUR (Gene\-ra\-litat de Catalunya), contract 2005SGR-00790, and by
RFBR, project 06-01-00609.

\end{document}